\documentstyle[multicol,aps,psfig]{revtex}
\renewcommand{\narrowtext}{\begin{multicols}{2}
\global\columnwidth20.5pc\noindent}
\renewcommand{\widetext}{\end{multicols}
\global\columnwidth42.5pc}
\begin{document}
\draft
\preprint{28 April 1999}
\title{Broken-Symmetry Ground States of Halogen-Bridged Binuclear
       Metal Complexes}
\author{Shoji Yamamoto}
\address
{Department of Physics, Okayama University,
 Tsushima, Okayama 700-8530, Japan}
\date{28 April 1999}
\maketitle
\begin{abstract}
Based on a symmetry argument, we study ground states of what we call
MMX-chain compounds, which are the new class of halogen-bridged metal
complexes.
Commensurate density-wave solutions of a relevant multi-band
Peierls-Hubbard model are systematically revealed within the
Hartree-Fock approximation.
We numerically draw ground-state phase diagrams, where various novel
density-wave states appear.
\end{abstract}
\narrowtext

   The competition between electron-electron (el-el) and
electron-phonon (el-ph) interactions in one dimension is a
fascinating subject to be studied.
Halogen-bridged transition-metal linear-chain complexes (MX chains)
are good target materials in this context and have a long history of
both chemical and physical researches.
The mixed-valence MX chains have been attracting extensive interest
primarily due to their significant dichroism and unusual Raman
spectra \cite{Clar95}, while the discovery of the mono-valence MX
chains \cite{Toft61,Tori41,Okam81}, whose ground states are Mott
insulators instead of Peierls insulators, aroused a renewed interest
in this system.
These materials, showing the variety of their transition metals,
ligands, bridging halogens, and counter ions, enable us to
investigate systematically quasi-one-dimensional electronic states
\cite{Okam09}.

   One of the most interesting consequences of intrinsic multi-band
effects and competing el-el and el-ph interactions is the variety of
ground states.
Employing a single-band model discarding the X$p_z$ orbitals,
Nasu \cite{Nasu65} visualized highly competing one-dimensinal ground
states.
Extensive and systematic two-band-model study was presented by the
Los Alamos group \cite{Gamm08,Bati28,Rode98,Bish03}.
Their vigorous study, covering even incommensurate density waves,
revealed various novel phases.
Their unique suggestion \cite{Gamm77} is also worth mentioning
that the MX-chain system can be regarded as a one-dimensional
analog of the CuO$_2$ plane in high-$T_{\rm c}$ superconductors.

   In an attempt to explore further into the family of these
materials, new halogen-bridged metal complexes, which are abbreviated
as MMX chains, have been synthesized, where binuclear metal units are
bridged by halogen ions.
K$_4$[Pt$_2$(P$_2$O$_5$H$_2$)$_4$X]$\cdot$$n$H$_2$O
(${\rm X}={\rm Cl},{\rm Br},{\rm I}$)
\cite{Che04,Kurm20,Clar09,Butl55} and
M$_2$(CH$_3$CS$_2$)$_4$I (${\rm M}={\rm Pt},{\rm Ni}$)
\cite{Bell44,Bell15} have been attracting considerable attention in
the chemical field, whereas they have less been studied in the
physical field \cite{Kimu40,Kita} so far.
The MMX-chain system may exhibit a wider variety of electronic
structures than the MX-chain system.
In this letter, focusing on commensurate density-wave ground
states, we present a systematic symmetry argument
\cite{Ozak55,Yama29,Yama22} and clarify what kinds of broken-symmetry
solutions are possible and actually stabilized within the
Hartree-Fock (HF) approximation.

   In order to treat both M- and X-atom electronic orbitals
explicitly, we employ the following one-dimensional model Hamiltonian
\widetext
\begin{eqnarray}
   {\cal H}
   &=& \sum_{n,s}
      \left[
       (\varepsilon_{\rm M}-\beta u_{3:n})n_{1:n,s}
      +(\varepsilon_{\rm M}+\beta u_{3:n})n_{2:n,s}
      + \varepsilon_{\rm X}               n_{3:n,s}
      \right]
    + \sum_{n}
      Ku_{3:n}^2
      \nonumber \\
   &-&\sum_{n,s}
      \left[
       (t_{\rm MX}-\alpha u_{3:n})a_{1:n,s}^\dagger a_{3:n,s}
      +(t_{\rm MX}+\alpha u_{3:n})a_{2:n,s}^\dagger a_{3:n,s}
      + t_{\rm MM}\,a_{1:n,s}^\dagger a_{2:n-1,s}
      +{\rm H.c.}
      \right]
      \nonumber \\
   &+&\sum_{n}
      (U_{\rm M}\,n_{1:n,+}n_{1:n,-}
      +U_{\rm M}\,n_{2:n,+}n_{2:n,-}
      +U_{\rm X}\,n_{3:n,+}n_{3:n,-})
      \nonumber \\
   &+&\sum_{n,s,s'}
      (V_{\rm MX}\,n_{1:n,s}n_{3:n  ,s'}
      +V_{\rm MX}\,n_{2:n,s}n_{3:n  ,s'}
      +V_{\rm MM}\,n_{1:n,s}n_{2:n-1,s'}) \,,
   \label{E:Hn}
\end{eqnarray}
\narrowtext
where $n_{i:n,s}=a_{i:n,s}^\dagger a_{i:n,s}$ with
$a_{1:n,s}^\dagger$, $a_{2:n,s}^\dagger$ and $a_{3:n,s}^\dagger$
being the creation operators of an electron with spin $s=\pm$ (up
and down) in the M$d_{z^2}$ and X$p_z$ orbitals at the $n$th MXM
unit, respectively, and $u_{3:n}$ the chain-direction displacement of
the halogen ion from the midpoint between the metal ions at the $n$th
MXM unit.
Here we have assumed that the lattice distortion comes only from the
halogen ions, which was the case with MX chains with ligand-locked
metal ions.
Although metal ions still look locked in the surrounding ligands in
the present system, there seems to be an experiment \cite{Kita}
implying movable metal ions in certain MMX chains.
However, the present model normally has plenty of parameters to be
controlled and therefore we assume fixed metal ions in our first
attempt.
The momentum representation of the Hamiltonian is generally given by
\widetext
\begin{eqnarray}
   {\cal H}
   &=& \sum_{i,j}\sum_{k,q}\sum_{s}
       \langle i:k+q|t|j:k\rangle
       a_{i:k+q,s}^\dagger a_{j:k,s}
    +  K\sum_k u_{3:k}u_{3:k}^*
       \nonumber \\
   &+& \frac{1}{2}\sum_{i,j,m,n}\sum_{k,k',q}\sum_{s,t}
       \langle i:k+q;m:k'|v|j:k;n:k'+q\rangle
       a_{i:k+q,s}^\dagger a_{m:k',t}^\dagger
       a_{n:k'+q,t} a_{j:k,s} \,,
   \label{E:Hk}
\end{eqnarray}
\narrowtext
where
the spin-free one-body and two-body interactions,
$\langle i:k+q|t|j:k\rangle$ and
$\langle i:k+q;m:k'|v|j:k;n:k'+q\rangle$,
are straightforwardly obtained from the Hamiltonian (\ref{E:Hn}).

   Let us introduce the symmetry group of the system as
\begin{equation}
   {\bf G}={\bf P}\times{\bf S}\times{\bf T} \,,
   \label{E:G}
\end{equation}
where ${\bf P}={\bf L}_1\land{\bf C}_2$ is the space group of a linear
chain with the one-dimensional translation group ${\bf L}_1$ whose
basis vector is the unit-cell translation $l_1$,
${\bf S}$ the group of spin rotation, and
${\bf T}$ the group of time reversal.
Here we have discarded the gauge group arriving at superconducting
phases.
Group actions on the electron operators are
defined as follows \cite{Ozak55}:
\begin{eqnarray}
   &&
   l\in{\bf L}_1:\ \ \ \ \ \ \ \ 
   l\cdot a_{i:k,s}^\dagger={\rm e}^{-{\rm i}kl}a_{i:k,s}^\dagger
   \,,\label{E:GAl} \\
   &&
   p\in{\bf C}_2:\ \ \ \ \ \ \,
   p\cdot a_{i:k,s}^\dagger=a_{i:pk,s}^\dagger
   \,,\label{E:GAp} \\
   &&
   u(\mbox{\boldmath$e$},\theta)\in{\bf S}:\ 
   u(\mbox{\boldmath$e$},\theta)\cdot a_{i:k,s}^\dagger
      =\sum_{s'}
       [u(\mbox{\boldmath$e$},\theta)]_{ss'} a_{i:k,s'}^\dagger
   \,,\label{E:GAu} \\
   &&
   t\in{\bf T}:\ \ \ \ \ \ \ \ \,
   t\cdot(f a_{i:k,s}^\dagger)
      =-s\,f^* a_{i:-k,-s}^\dagger
   \,,\label{E:GAt}
\end{eqnarray}
where
$f$ is an arbitrary complex number.
The spin rotation of angle $\theta$ around an axis
$\mbox{\boldmath$e$}$,
$u(\mbox{\boldmath$e$},\theta)$,
is explicitly represented as
$\sigma^0\cos(\theta/2)
 -(\mbox{\boldmath$\sigma$}\cdot\mbox{\boldmath$e$})\sin(\theta/2)$
in terms of the $2\times 2$ unit matrix $\sigma^0$ and
a vector composed of the Pauli-matrices,
$\mbox{\boldmath$\sigma$}=(\sigma^x,\sigma^y,\sigma^z)$.

   Let $\check{G}$ denote the irreducible representations of {\bf G}
over the real number field, where their representation space is spanned
by the Hermitian operators \{$a_{i:k,s}^\dagger a_{j:k',s'}$\}.
There is a one-to-one correspondence between $\check{G}$ and
broken-symmetry phases of density-wave type.
Any representation $\check{G}$ is obtained as a Kronecker product of
the irreducible real representations of ${\bf P}$, ${\bf S}$, and
${\bf T}$:
\begin{equation}
   \check{G}=\check{P}\otimes\check{S}\otimes\check{T} \,.
   \label{E:Grep}
\end{equation}
$\check{P}$ is characterized by an ordering vector $q$ in the
Brillouin zone and an irreducible representation of its little group
${\bf P}(q)$, and is therefore labeled $q\check{P}(q)$.
The relevant representations of ${\bf S}$ and ${\bf T}$ are,
respectively, given by
\begin{eqnarray}
   &&
   \check{S}^0(u(\mbox{\boldmath$e$},\theta))
      =1 \,,\ \ 
   \check{S}^1(u(\mbox{\boldmath$e$},\theta))
      =O(u(\mbox{\boldmath$e$},\theta)) \,,
   \label{E:Srep} \\
   &&
   \check{T}^0(t)=1\,,\ \ 
   \check{T}^1(t)=-1\,,
   \label{E:Trep}
\end{eqnarray}
where $O(u(\mbox{\boldmath$e$},\theta))$ is the $3\times 3$
orthogonal matrix satisfying
$
   u(\mbox{\boldmath$e$},\theta)
   \mbox{\boldmath$\sigma$}^\lambda
   u^\dagger(\mbox{\boldmath$e$},\theta)
      =\sum_{\mu=x,y,z}
       [O(u(\mbox{\boldmath$e$},\theta))]_{\lambda\mu}
       \mbox{\boldmath$\sigma$}^\mu \ \
       (\lambda=x,\,y,\,z)
$.
The representations
$\check{P}\otimes\check{S}^0\otimes\check{T}^0$,
$\check{P}\otimes\check{S}^1\otimes\check{T}^1$,
$\check{P}\otimes\check{S}^0\otimes\check{T}^1$, and 
$\check{P}\otimes\check{S}^1\otimes\check{T}^0$
correspond to charge-density-wave, spin-density-wave,
charge-current-wave, and spin-current-wave states, respectively.
We leave out current-wave states, because in one dimension all
of them but one-way uniform-current states break the charge- or
spin-conservation law.
We consider two ordering vectors $q=0$ and $q=\pi$, which are labeled
$\mit\Gamma$ and $X$, respectively.
Thus the instabilities labeled
$K\check{P}(K)
 \otimes\check{S}^i\otimes\check{T}^i$
($K={\mit\Gamma},X$; $i=0,1$) are of our interest.
Since ${\bf P}({\mit\Gamma})={\bf P}(X)={\bf C}_2$,
$\check{P}({\mit\Gamma})$ and $\check{P}(X)$ are either
$A$ ($C_2$-symmetric) or $B$ ($C_2$-antisymmetric) representation of
${\bf C}_2$.

   In the HF approximation the Hamiltonian (\ref{E:Hk}) is replaced
by
\widetext
\begin{equation}
   {\cal H}_{\rm HF}
      =\sum_{i,j}\sum_{k,s,s'}\sum_{\lambda=0,z}
       \left[
         x_{ij}^{\lambda}({\mit\Gamma};k)
         a_{i:k,s}^\dagger a_{j:k,s'}
        +x_{ij}^{\lambda}(X;k)
         a_{i:k+\pi,s}^\dagger a_{j:k,s'}
       \right]
       \sigma_{ss'}^\lambda\,.
   \label{E:HHF}
\end{equation}
The present model arrives at no helical-spin ($\lambda=x,y$) solution.
The self-consistent field $x_{ij}^\lambda(K;k)$ is expressed as
\begin{eqnarray}
   &&
   x_{ij}^0({\mit\Gamma};k)
     =\langle i:k|t|j:k \rangle
     +\sum_{m,n}\sum_{k'}
      \rho_{nm}^0({\mit\Gamma};k')
   \nonumber \\
   &&
   \quad\times
   \left(
    2\langle i:k;m:k'|v|j:k ;n:k' \rangle
    -\langle i:k;m:k'|v|n:k';j:k  \rangle
   \right)\,,
   \\
   &&
   x_{ij}^0(X;k)
     =\langle i:k+\pi|t|j:k \rangle
     +\sum_{m,n}\sum_{k'}
      \rho_{nm}^0(X;k'+\pi)
   \nonumber \\
   &&
   \quad\times
   \left(
    2\langle i:k+\pi;m:k'|v|j:k     ;n:k'+\pi \rangle
    -\langle i:k+\pi;m:k'|v|n:k'+\pi;j:k      \rangle
   \right)\,,
   \\
   &&
   x_{ij}^z({\mit\Gamma};k)
     =-\sum_{m,n}\sum_{k'}
      \rho_{nm}^z({\mit\Gamma};k')
      \langle i:k;m:k'|v|n:k';j:k \rangle\,,
   \\
   &&
   x_{ij}^z(X;k)
     =-\sum_{m,n}\sum_{k'}
          \rho_{nm}^z(X;k'+\pi)
          \langle i:k+\pi;m:k'\vert v\vert n:k'+\pi;j:k\rangle\,,
   \label{E:SCF}
\end{eqnarray}
\narrowtext
in terms of the density matrices
\begin{equation}
   \left.
   \begin{array}{lll}
   \rho_{ij}^\lambda({\mit\Gamma};k)
   &=& {\displaystyle\frac{1}{2}\sum_{s,s'}}
       \langle a_{j:k,s}^\dagger a_{i:k,s'}\rangle_{\rm HF}\,
       \sigma_{ss'}^\lambda \,,
   \\
   \rho_{ij}^\lambda(X;k)
   &=& {\displaystyle\frac{1}{2}\sum_{s,s'}}
       \langle a_{j:k+\pi,s}^\dagger a_{i:k,s'}\rangle_{\rm HF}\,
       \sigma_{ss'}^\lambda \,,
   \end{array}
   \right.
   \label{E:rho}
\end{equation}
where $\langle\cdots\rangle_{\rm HF}$ means the quantum average in
a HF eigenstate.
The HF Hamiltonian (\ref{E:HHF}) is decomposed as \cite{Ozak14}
\begin{equation}
  {\cal H}_{\rm HF}
   =\sum_{D=A,B}\sum_{K={\mit\Gamma},X}\sum_{\lambda=0,z}
    h^\lambda(K;D)\,,
\end{equation}
where the irreducible components $h^\lambda(K;D)$ are given by
\begin{equation}
 h^\lambda(K;D)
 =\frac{1}{2}\sum_{p\in{\bf C}_2}
  \chi^{(D)}(p)\,p\cdot
  x_{ij}^\lambda(K;k)a_{i:k,s}^\dagger a_{j:k,s'}
  \sigma_{ss'}^\lambda\,,
\end{equation}
with $\chi^{(D)}(p)$ being the irreducible character of the $D$
representation for the group element $p$.
Now we obtain the relevant broken-symmetry Hamiltonian for the
representation $KD\otimes\check{S}^i\otimes\check{T}^i$ as
$h^0({\mit\Gamma};A)+h^\lambda(K;D)$, where $\lambda=0$ for $i=0$ and
$\lambda=z$ for $i=1$.

\begin{figure}
\vskip 2mm
\mbox{\psfig{figure=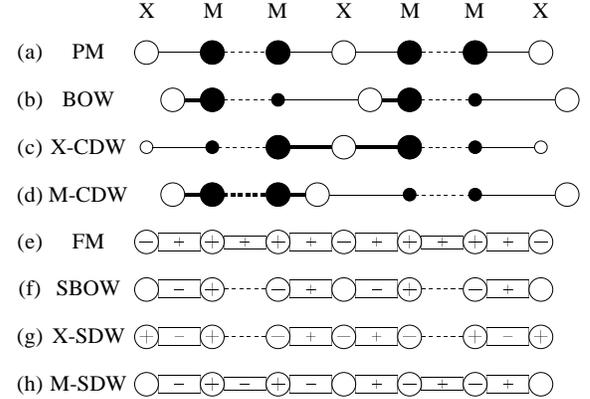,width=80mm,angle=0}}
\vskip 5mm
\caption{Schematic representation of possible density-wave states,
         where the variety of circles and segments qualitatively
         represents the variation of charge densities and bond
         orders, respectively, whereas the signs $\pm$ in circles
         and strips describe the alternation of spin densities and
         spin bond orders, respectively.
         Unpainted circles shifted from the midpoint between the
         neighboring painted circles qualitatively represent
         halogen-ion displacements.}
\label{F:DW}
\end{figure}

   The charge and spin densities on site $i$ at the $n$th MXM unit
are, respectively, expressed as
\begin{eqnarray}
   d_{i:n}
     &=&\sum_s\langle a_{i:n,s}^\dagger a_{i:n,s}\rangle_{\rm HF}
   \,, \label{E:OPd} \\
   s_{i:n}^z
     &=&\frac{1}{2}\sum_{s,s'}
     \langle a_{i:n,s}^\dagger a_{i:n,s'}\rangle_{\rm HF}
     \sigma_{ss'}^z
   \,, \label{E:OPs}
\end{eqnarray}
while the bond and spin bond orders between site $i$ at the $n$th MXM
unit and site $j$ at the $m$th MXM unit are, respectively, defined as
\begin{eqnarray}
   p_{i:n;j:m}
     &=&\sum_s
     \langle a_{i:n,s}^\dagger a_{j:m,s}\rangle_{\rm HF}
   \,, \label{E:OPp} \\
   t_{i:n;j:m}^z
     &=&\frac{1}{2}\sum_{s,s'}
     \langle a_{i:n,s}^\dagger a_{j:n,s'}\rangle_{\rm HF}
     \sigma_{ss'}^z
   \,. \label{E:OPt}
\end{eqnarray}
The halogen-ion displacements $u_n$ are self-consistently determined
so as to minimize the HF energy 
$E_{\rm HF}=\langle{\cal H}\rangle_{\rm HF}$.
All the order parameters (\ref{E:OPd})-(\ref{E:OPt}), as well as
$E_{\rm HF}$, are expressed in terms of the density matrices whose
symmetry properties are definitely determined by the corresponding
invariance group (Table \ref{T:IG}).
Here we simply describe all the phases obtained and schematically
show them in Fig. \ref{F:DW}.
\vspace{-0.5mm}
\begin{description}
\item[\rm\quad(a)]${\mit\Gamma}A\otimes\check{S}^0\otimes\check{T}^0$
\vspace{-1.0mm}
\item[]\qquad The paramagnetic state with the full symmetry
       (\ref{E:G}), abbreviated as PM.
\vspace{-1.0mm}
\item[\rm\quad(b)]${\mit\Gamma}B\otimes\check{S}^0\otimes\check{T}^0$
\vspace{-1.0mm}
\item[]\qquad Electron-lattice-coupled bond order wave accompanied
       by alternating metal charge densities, abbreviated as BOW.
\vspace{-1.0mm}
\item[\rm\quad(c)]$XA           \otimes\check{S}^0\otimes\check{T}^0$
\vspace{-1.0mm}
\item[]\qquad Charge density wave on halogen sites accompanied by
       alternating metal charge densities, abbreviated as X-CDW.
\vspace{-1.0mm}
\item[\rm\quad(d)]$XB           \otimes\check{S}^0\otimes\check{T}^0$
\vspace{-1.0mm}
\item[]\qquad Electron-lattice-coupled charge density wave on metal
       sites, abbreviated as M-CDW.
\vspace{-1.0mm}
\item[\rm\quad(e)]${\mit\Gamma}A\otimes\check{S}^1\otimes\check{T}^1$
\vspace{-1.0mm}
\item[]\qquad Ferromagnetism with uniform spin bond orders,
       abbreviated as FM.
\vspace{-1.0mm}
\item[\rm\quad(f)]${\mit\Gamma}B\otimes\check{S}^1\otimes\check{T}^1$
\vspace{-1.0mm}
\item[]\qquad Spin bond order wave accompanied by alternating metal
       spin densities, abbreviated as SBOW.
\vspace{-1.0mm}
\item[\rm\quad(g)]$XA           \otimes\check{S}^1\otimes\check{T}^1$
\vspace{-1.0mm}
\item[]\qquad Spin density wave on halogen sites accompanied by
       alternating spin bond orders and metal spin densities,
       abbreviated as X-SDW.
\vspace{-1.0mm}
\item[\rm\quad(h)]$XB           \otimes\check{S}^1\otimes\check{T}^1$
\vspace{-1.0mm}
\item[]\qquad Spin density wave on metal sites accompanied by
       alternating spin bond orders, abbreviated as M-SDW.
\vspace{-0.5mm}
\end{description}
Magnetic instabilities are generally not coupled with phonons.
The reason X-CDW is a purely electronic state is just because we have
restricted the lattice distortion to halogen ions.
X-CDW is not stabilized within our assumption but may be by metal-ion
displacements.

\begin{figure}
\mbox{\psfig{figure=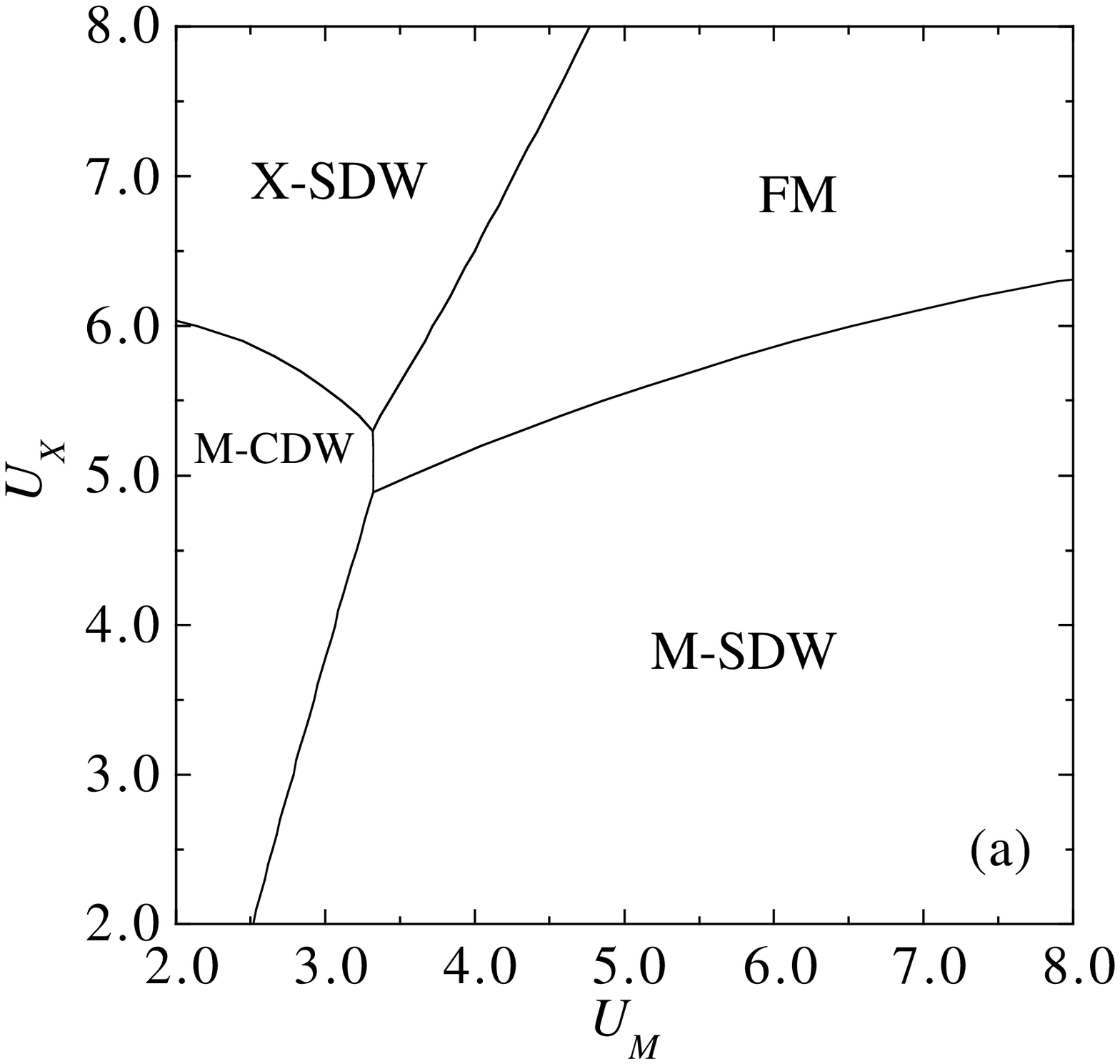,width=70mm,angle=0}}
\vskip -24mm
\mbox{\psfig{figure=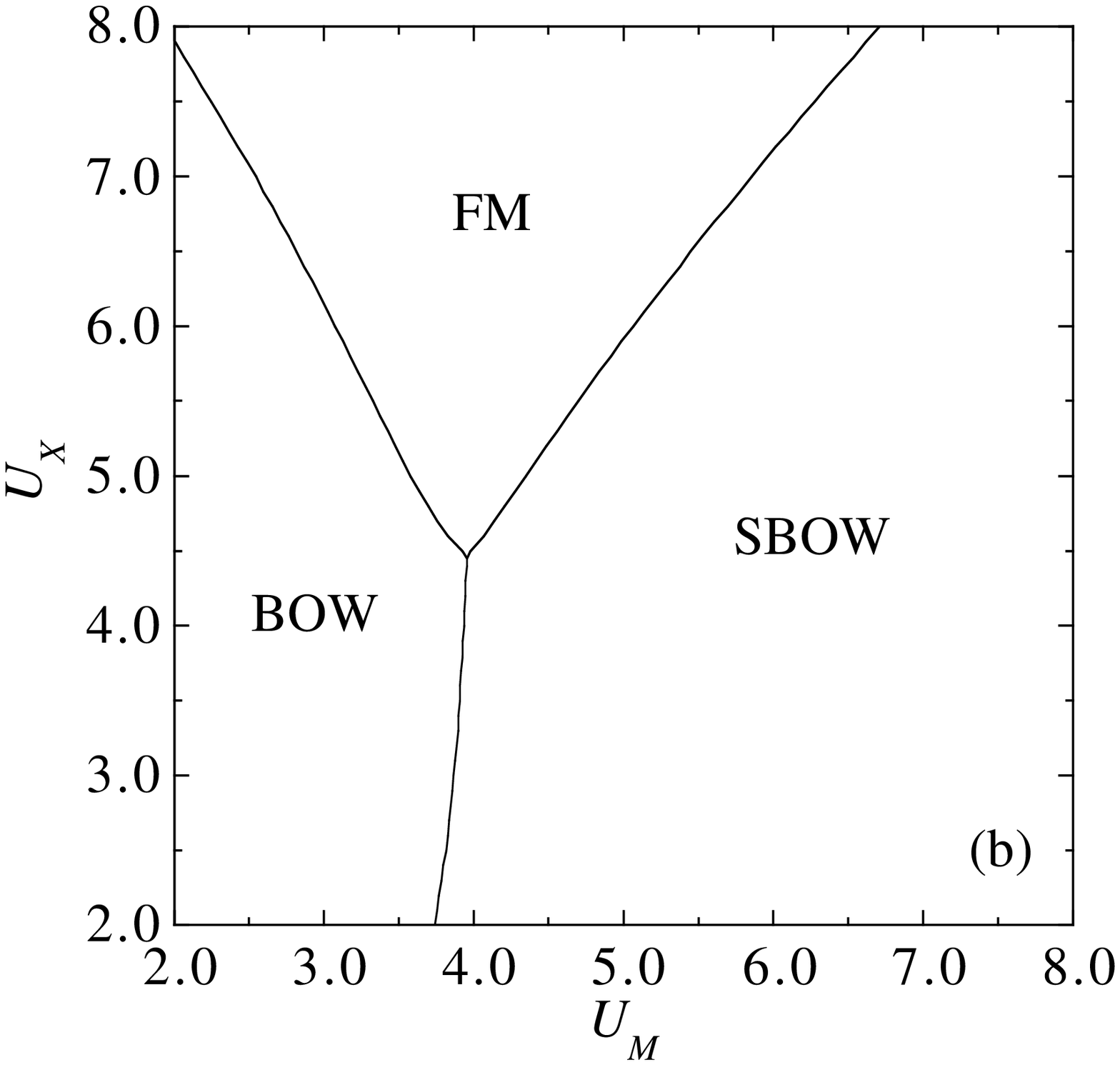,width=70mm,angle=0}}
\vskip -24mm
\mbox{\psfig{figure=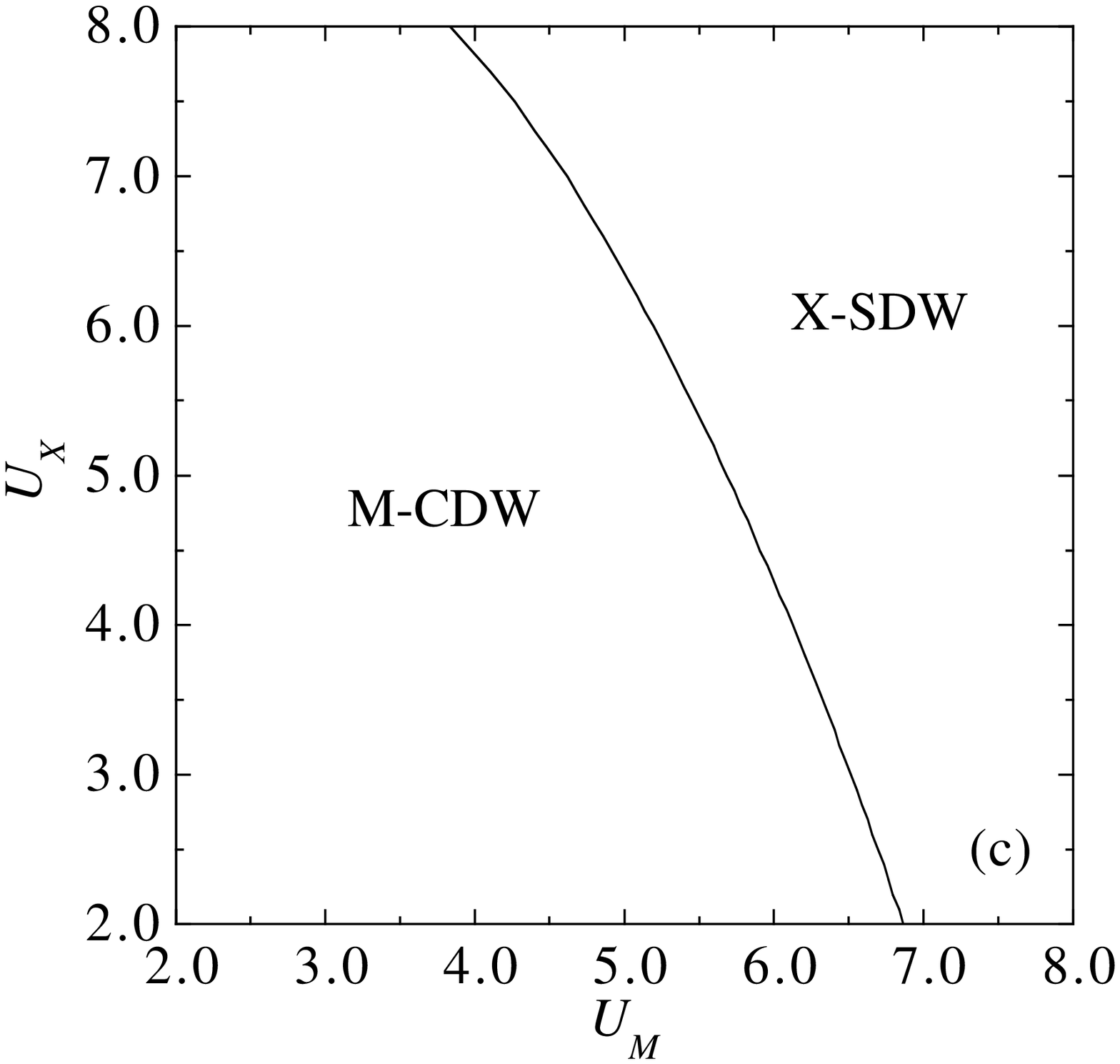,width=70mm,angle=0}}
\vskip -22mm
\caption{Typical ground-state phase diagrams at various band
         fillings:
         (a) $\frac{5}{6}$ band filling, $\alpha=1.2$, $\beta=0.6$;
         (b) $\frac{4}{6}$ band filling, $\alpha=1.0$, $\beta=0.5$;
         (c) $\frac{3}{6}$ band filling, $\alpha=1.0$, $\beta=0.5$.
         Here the following parametrization is common to all the
         cases:
         $\varepsilon_{\rm M}-\varepsilon_{\rm X}=1.0$,
         $V_{\rm MM}=1.0$, $V_{\rm MX}=1.0$,
         and $t_{\rm MM}$, $t_{\rm MX}$, and $K$ are all set to
         unity.}
\label{F:PhD}
\end{figure}

   Let us observe a few pieces of numerical investigation.
We show in Fig. \ref{F:PhD} typical ground-state phase diagrams
obtained by computing $E_{\rm HF}$.
As-grown MMX-chain compounds have the $\frac{5}{6}$-filled band
structure, where $q=\pi$ instabilities are dominant unless both
$U_{\rm M}$ and $U_{\rm X}$ are large enough.
Once we go beyond the mean-field theory, the ferromagnetic region
should more or less shrink due to multiscattering effects
\cite{Kana75}.
As holes are doped into the system, $q=\pi$ instabilities are
generally replaced by $q=0$ ones.
Further hole doping beyond the $\frac{4}{6}$ band filling again
stabilizes $q=\pi$ phases.
However, in the heavily hole-doped region where the closed-shell
electronic structure of halogen ions is broken, M-SDW with no spin
density on halogen sites seems to be much less stabilized.
Energy calculations suggest that all the transitions between
broken-symmetry phases are of first order.
Further numerical investigations, including microscopic information
on the electronic structure, will be presented elsewhere.
It seems that physical research on MMX-chain compounds is still in
its early stage.
We hope the present calculation will motivate and accelerate further
synthesis and measurements of these fascinating materials.

   It is a pleasure to thank H. Okamoto, M. Kuwabara, and K.
Yonemitsu for helpful comments and fruitful discussions.
This work was supported by the Japanese Ministry of Education,
Science, and Culture and the Sanyo-Broadcasting Foundation for
Science and Culture.
The numerical calculation was done using the facility of the
Supercomputer Center, Institute for Solid State Physics, University
of Tokyo.

\begin{table}
\caption{Invariance groups characteristic of each irreducible
         representation.}
\begin{tabular}{cl}
Representation & Invariance group \\
\tableline
\noalign{\vskip 2pt}
${\mit\Gamma}A\otimes\check{S}^0\otimes\check{T}^0$ &
${\bf L}_1{\bf C}_2{\bf S}{\bf T}$                  \\
${\mit\Gamma}B\otimes\check{S}^0\otimes\check{T}^0$ &
${\bf L}_1{\bf S}{\bf T}$                           \\
$X           A\otimes\check{S}^0\otimes\check{T}^0$ &
${\bf L}_2{\bf C}_2{\bf S}{\bf T}$                  \\
$X           B\otimes\check{S}^0\otimes\check{T}^0$ &
$(1+l_1C_2){\bf L}_2{\bf S}{\bf T}$                 \\
${\mit\Gamma}A\otimes\check{S}^1\otimes\check{T}^1$ &
${\bf L}_1{\bf C}_2
 {\bf A}(\mbox{\boldmath$e$}_z)
 {\bf M}(\mbox{\boldmath$e$}_\parallel)$            \\
${\mit\Gamma}B\otimes\check{S}^1\otimes\check{T}^1$ &
$(1+C_2u(\mbox{\boldmath$e$}_\parallel,\pi))
 {\bf L}_1
 {\bf A}(\mbox{\boldmath$e$}_z)
 {\bf M}(\mbox{\boldmath$e$}_\parallel)$            \\
$X           A\otimes\check{S}^1\otimes\check{T}^1$ &
$(1+l_1u(\mbox{\boldmath$e$}_\parallel,\pi))
 {\bf L}_2{\bf C}_2
 {\bf A}(\mbox{\boldmath$e$}_z)
 {\bf M}(\mbox{\boldmath$e$}_\parallel)$            \\
$X           B\otimes\check{S}^1\otimes\check{T}^1$ &
$(1+l_1C_2)
 (1+l_1u(\mbox{\boldmath$e$}_\parallel,\pi))
 {\bf L}_2
 {\bf A}(\mbox{\boldmath$e$}_z)
 {\bf M}(\mbox{\boldmath$e$}_\parallel)$            \\
\end{tabular}
${\bf L}_2=\{E,l_1^{2n}|\,n\in\mbox{\boldmath$N$}\}$.
\hfill\break
${\bf A}(\mbox{\boldmath$e$}_z)
 =\{u(\mbox{\boldmath$e$}_z,\theta)
  |\,0\leq\theta\leq 4\pi,
   \mbox{\boldmath$e$}_z\|(\mbox{the $z$ direction})\}$.
\hfill\break
${\bf M}(\mbox{\boldmath$e$}_\parallel)
 =\{E,tu(\mbox{\boldmath$e$}_\parallel,\pi)
  |\,\mbox{\boldmath$e$}_\|\|(\mbox{the chain direction})\}$.
\label{T:IG}
\end{table}

\widetext

\begin{references}
\bibitem{Clar95}
   R. J. H. Clark,
      in {\it Advances in Infrared and Raman Spectroscopy},
      edited by R. J. H. Clark and R. E. Hester
      (Wiley, New York, 1984) Vol. II, p. 95.
\bibitem{Toft61}
   H. Toftlund and O. Simonsen,
      Inorg. Chem. {\bf 23}, 4261 (1984).

\bibitem{Tori41}
   K. Toriumi, Y. Wada, T. Mitani, S. Bandow, M. Yamashita, and
   Y. Fujii,
      J. Am. Chem. Soc. {\bf 111}, 2341 (1989).

\bibitem{Okam81}
   H. Okamoto, T. Mitani, K. Toriumi, and M. Yamashita,
      Phys. Rev. B {\bf 42}, 10381 (1990).

\bibitem{Okam09}
   H. Okamoto, T. Mitani, K. Toriumi, and M. Yamashita,
      Mater. Sci. Eng. B {\bf 13}, L9 (1992).

\bibitem{Nasu65}
   K. Nasu,
      J. Phys. Soc. Jpn. {\bf 52}, 3865 (1983);
                         {\bf 53},  302 (1984).

\bibitem{Gamm08}
   J. T. Gammel, A. Saxena, I. Batisti\'c, A. R. Bishop, and
   S. R. Phillpot,
      Phys. Rev. B {\bf 45}, 6408 (1992).

\bibitem{Bati28}
   I. Batisti\'c, J. T. Gammel, and A. R. Bishop,
      Phys. Rev. B {\bf 44}, 13228 (1991).

\bibitem{Rode98}
   H. R\"oder, A. R. Bishop, and J. T. Gammel,
      Phys. Rev. Lett. {\bf 70}, 3498 (1993).

\bibitem{Bish03}
   A. R. Bishop,
      Synth. Met. {\bf 86}, 2203 (1997).

\bibitem{Gamm77}
   J. T. Gammel, K. Yonemitsu, A. Saxena, A. R. Bishop, and
   H. R\"oder,
      Synth. Met. {\bf 55}-{\bf 57}, 3377 (1993).

\bibitem{Che04}
   C.-M. Che, F. H. Herbstein, W. P. Schaefer, R. E. Marsh, and
   H. B. Gray,
      J. Am. Chem. Soc. {\bf 105}, 4604 (1983).

\bibitem{Kurm20}
   M. Kurmoo and R. J. H. Clark,
      Inorg. Chem. {\bf 24}, 4420 (1985).

\bibitem{Clar09}
   R. J. H. Clark, M. Kurmoo, H. M. Dawes, and M. B. Hursthouse,
      Inrog. Chem. {\bf 25}, 409 (1986).

\bibitem{Butl55}
   L. G. Butler, M. H. Zietlow, C.-M. Che, W. P. Schaefer,
   S. Sridhar, P. J. Grunthaner, B. I. Swanson, R. J. H. Clark, and
   H. B. Gray,
      J. Am. Chem. Soc. {\bf 110}, 1155 (1988).

\bibitem{Bell44}
   C. Bellitto, A. Flamini, L. Gastaldi, and L. Scaramuzza,
      Inorg. Chem. {\bf 22}, 444 (1983).

\bibitem{Bell15}
   C. Bellitto, G. Dessy, and V. Fares,
      Inorg. Chem. {\bf 24}, 2815 (1985).

\bibitem{Kimu40}
   N. Kimura, H. Ohki, R. Ikeda, M. Yamashita,
      Chem. Phys. Lett. {\bf 220}, 40 (1994).

\bibitem{Kita}
   H. Kitagawa, T. Sonoyama, M. Yamamoto, T. Mitani, K. Toriumi, and
   M. Yamashita,
      to be published in Synth. Met. (1999).

\bibitem{Ozak55}
   M. Ozaki,
      Int. J. Quantum Chem. {\bf 42}, 55 (1992);
   S. Yamamoto and M. Ozaki, {\it ibid.} {\bf 44}, 949 (1992).

\bibitem{Yama29}
   S. Yamamoto and M. Ozaki,
      Solid State Commun. {\bf 83}, 329 (1992);
                          {\bf 83}, 335 (1992).

\bibitem{Yama22}
   S. Yamamoto,
      Phys. Lett. A {\bf 247}, 422 (1998).

\bibitem{Ozak14}
   M. Ozaki,
      J. Math. Phys. {\bf 26}, 1514 (1985).

\bibitem{Kana75}
   J. Kanamori,
      Prog. Theor. Phys. {\bf 30}, 275 (1963).

\end{references}
\end{document}